\documentclass[fleqn,usenatbib]{mnras}

\usepackage[utf8]{inputenc}

\usepackage{acronym}
\usepackage{ae,aecompl}
\usepackage{booktabs}
\usepackage{dcolumn}
\usepackage{graphicx}
\usepackage{amsmath,amsfonts,amssymb}
\usepackage{mathrsfs}
\usepackage[normalem]{ulem}

\usepackage[T1]{fontenc}

\let\oldhref\href
\renewcommand{\href}[2]{\oldhref{#1}{\hbox{#2}}}

\newcolumntype{d}[1]{D{.}{.}{#1}}

\usepackage{color}

\newacro{ADM}{Arnowitt-Deser-Misner}
\newacro{AMR}{adaptive mesh-refinement}
\newacro{BH}{black hole}
\newacro{BBH}{binary black-hole}
\newacro{BHNS}{black-hole neutron-star}
\newacro{BNS}{binary neutron star}
\newacro{CCSN}{core-collapse supernova}
\newacroplural{CCSN}[CCSNe]{core-collapse supernovae}
\newacro{CMA}{consistent multi-fluid advection}
\newacro{DG}{discontinuous Galerkin}
\newacro{HMNS}{hypermassive neutron star}
\newacro{EM}{electromagnetic}
\newacro{ET}{Einstein Telescope}
\newacro{EOB}{effective-one-body}
\newacro{EOS}{equation of state}
\newacroplural{EOS}[EOSs]{equations of state}
\newacro{FF}{fitting factor}
\newacro{GR}{general relativity}
\newacro{GRLES}{general-relativistic large-eddy simulation}
\newacro{GRHD}{general-relativistic hydrodynamics}
\newacro{GRMHD}{general-relativistic magnetohydrodynamics}
\newacro{GW}{gravitational wave}
\newacro{ILES}{implicit large-eddy simulations}
\newacro{LIA}{linear interaction analysis}
\newacro{LES}{large-eddy simulation}
\newacroplural{LES}[LES]{large-eddy simulations}
\newacro{MNS}{massive neutron star}
\newacro{MRI}{magnetorotational instability}
\newacro{NR}{numerical relativity}
\newacro{NS}{neutron star}
\newacro{PNS}{protoneutron star}
\newacro{SASI}{standing accretion shock instability}
\newacro{SGRB}{short $\gamma$-ray burst}
\newacro{SMNS}{supramassive neutron star}
\newacro{SN}{supernova}
\newacroplural{SN}[SNe]{supernovae}
\newacro{SNR}{signal-to-noise ratio}

\title{Long-lived Remnants from Binary Neutron Star Mergers}
\author[D.~Radice, A.~Perego, S.~Bernuzzi, and B.~Zhang]{
David Radice$^{1,2}$,
Albino Perego$^{3,4,5}$,
Sebastiano Bernuzzi$^{6,3,4}$, and
Bing Zhang$^{7}$ \\
$^1$ Institute for Advanced Study, 1 Einstein Drive,
Princeton, NJ 08540, USA \\
$^2$ Department of Astrophysical Sciences, Princeton University,
4 Ivy Lane, Princeton, NJ 08544, USA \\
$^3$ Dipartimento di Scienze Matematiche, Fisiche ed Informatiche,
Universit\'a di Parma, I-43124 Parma, Italy \\
$^4$ Istituto Nazionale di Fisica Nucleare, Sezione Milano Bicocca,
gruppo collegato di Parma, I-43124 Parma, Italy \\
$^5$ Dipartimento di Fisica, Universit\`{a} degli Studi di Milano Bicocca,
Piazza della Scienza 3, 20126 Milano, Italia \\
$^6$ Theoretisch-Physikalisches Institut,
Friedrich-Schiller-Universit\"at Jena, D-07743 Jena, Germany\\
$^7$ Department of Physics and Astronomy, University of Nevada Las
Vegas, NV 89154, USA
}

\begin{document}
\label{firstpage}
\pagerange{\pageref{firstpage}--\pageref{lastpage}}

\maketitle
\begin{abstract}
Massive neutron star (NS) with lifetimes of at least several seconds are
expected to be the result of a sizable fraction of NS mergers. We study
their formation using a large set of numerical relativity simulations.
We show that they are initially endowed with angular momentum that
significantly exceeds the mass-shedding limit for rigidly-rotating
equilibria. We find that gravitational-wave (GW) emission is not able to
remove this excess angular momentum within the time over which solid
body rotation should be achieved. Instead, we argue that the excess
angular momentum could be carried away by massive winds. Long-lived
merger remnants are also formed with larger gravitational masses than
those of rigidly-rotating NSs having the same number of baryons. The
excess mass is likely radiated in the form of neutrinos. The evolution
of long-lived remnants on the viscous timescale is thus determined by
the interplay of finite-temperature effects, mass ejection, and
neutrinos with potentially dramatic consequences for the remnants'
properties and stability. We also provide an empirical fit for the spin
of the remnant at the end of its viscous evolution as a function of its
final mass, and we discuss the implications for the magnetar model of
short gamma-ray bursts (SGRBs). Finally, we investigate the possible
electromagnetic signatures associated with the viscous ejecta. Massive
outflows possibly resulting from the formation of long-lived remnants
would power unusually bright, blue kilonova counterparts to GW events
and SGRBs whose detection would provide smoking gun evidence for the
formation of long-lived remnants.
\end{abstract}


\section{Introduction}

The outcome of \ac{NS} mergers depends on the total mass of the system
and on the poorly known \ac{EOS} of dense nuclear matter \citep[and
references therein]{shibata:2016book}. Binaries with mass larger than
${\sim}1.3{-}1.7$ times the maximum mass for a nonrotating \ac{NS}
result in prompt \ac{BH} formation \citep{hotokezaka:2011dh,
bauswein:2013jpa}. Binaries with lower masses, but above the maximum
mass of isolated rigidly rotating \acp{NS}, result in the formation of
\acp{HMNS} temporarily supported against gravitational collapse by the
large differential rotation \citep{baumgarte:1999cq, rosswog:2001fh,
shibata:2006nm, baiotti:2008ra, sekiguchi:2011zd, palenzuela:2015dqa,
bernuzzi:2015opx}. Even lower mass systems produce \ac{NS} remnants that
are stable on the spin down timescale (seconds to hours), called
supramassive NSs (SMNSs), or stable massive NSs (MNSs) if their mass is
below the maximum mass of a nonrotating \ac{NS}
\citep{hotokezaka:2013iia, giacomazzo:2013uua, foucart:2015gaa,
kastaun:2016yaf, ciolfi:2017uak, kiuchi:2017zzg}.
\acused{SMNS}\acused{MNS}

In the case of the binary \ac{NS} merger GW170817
\citep{theligoscientific:2017qsa, gbm:2017lvd, monitor:2017mdv}, the
most likely outcome was a \ac{HMNS} \citep{margalit:2017dij,
shibata:2017xdx, radice:2017lry}. However, the formation of a long-lived
remnant for GW170817 is not completely ruled out \citep{yu:2017syg,
ai:2018jtv, geng:2018vaa, li:2018hzy}. Indeed, the formation of
\acp{SMNS} is expected to occur in a sizable fraction of mergers
\citep{lasky:2013yaa, gao:2015xle, piro:2017zec}. This expectation has
been recently reinforced by the discovery of two double \ac{NS} systems
with total gravitational masses as low as $2.5\, M_\odot$
\citep{martinez:2017jbp, stovall:2018ouw}. Long-lived remnants have also
been invoked to explain late time X-rays excess seen in some short
gamma-ray bursts (SGRBs; \citealt{dai:1998bb, dai:1998hm, zhang:2000wx,
dai:2006hj, metzger:2007cd, rowlinson:2010a, bucciantini:2011kx,
rowlinson:2013ue, metzger:2013cha, rezzolla:2014nva, ciolfi:2014yla,
lu:2015rta, gao:2015xle, siegel:2015swa, siegel:2015twa,
margalit:2015qza, geng:2016noq, murase:2017snw}).  \acused{SGRB}

The identification of the outcome of the merger of binary \ac{NS}
systems with different masses would yield a precise measurement of the
maximum mass of \acp{NS} \cite[e.g.,][]{lasky:2013yaa, lawrence:2015oka,
piro:2017zec, margalit:2017dij, rezzolla:2017aly, ruiz:2017due,
drago:2018nzf, most:2018hfd}. This, in turn, would constrain the
\ac{EOS} of matter at extreme densities \citep{lattimer:2012nd}. It is
therefore important to identify signatures indicative of the formation
of long-lived remnants. The presence of temporarily extended X-ray
activity immediately after a merger would be one indication that a
\ac{BH} did not form in a dynamical timescale after the merger
\citep{metzger:2007cd, zhang:2012wt, sun:2016pcb, rowlinson:2013ue,
metzger:2013cha, siegel:2015swa, siegel:2015twa, gao:2016uwi,
wang:2016vvh, murase:2017snw}. Another would be the change in the
character of the optical counterpart to the merger due to the
irradiation of the ejecta by the central object \citep{metzger:2014ila,
lippuner:2017bfm}, the production of magnetized outflows
\citep{metzger:2018uni}, or the thermalization of the spin down
luminosity of the remnant \citep{yu:2013kra, metzger:2013cha,
gao:2015rua, siegel:2015swa, siegel:2015twa, kisaka:2015vma,
gao:2016uwi}. Finally, long-lived remnants could be revealed by the
appearance of bright radio flares raising on timescales of years from
the merger \citep{gao:2013rd, metzger:2013cka, gompertz:2014zwa,
hotokezaka:2015eja, horesh:2016dah, fong:2016orv}.

In this work, we employ general-relativistic merger simulations with
realistic microphysics to study the formation of long-lived remnants and
discuss their evolution during the subsequent viscous timescale. We show
that massive and supramassive \acp{NS} are born with angular momenta
significantly exceeding the mass-shedding limit for uniformly rotating
\acp{NS} and, as a consequence, they are likely to give rise to massive
outflows over the viscous timescale.  These could produce luminous
kilonova counterparts that would be smoking gun evidence for the
formation of massive or supramassive \acp{NS} if detected by future
UV/optical/infrared follow ups on \ac{GW} events or \acp{SGRB}. We also
constrain the spin of the remnants, and we discuss the implication of
our results for the magnetar model of \acp{SGRB} and the role of thermal
effects for the stability of the merger remnant. In our discussion
``remnant'' is used to indicate all gravitationally bound matter left
after the merger. Conversely, where needed, we use the expression
``\ac{NS} remnant'' to denote the centrally condensed part of the
remnant.

\section{Merger Remnants}
\label{sec: Merger Remnants}

\subsection{Simulation Setup}
Our analysis is based on the results of about 35 \ac{NS} merger
simulations performed with the \texttt{WhiskyTHC} code
\citep{radice:2012cu, radice:2013hxh, radice:2013xpa}. Our simulations
span a range of total gravitational masses $M_g = M_1 + M_2$ between
$2.4\, M_\odot$ and $3.4\, M_\odot$, and mass ratios $q = M_2/M_1$
between $0.85$ and $1.0$. We adopt 4 tabulated nuclear \ac{EOS} broadly
consistent with current laboratory and astrophysical constraints: the
DD2 EOS \citep{typel:2009sy, hempel:2009mc}, the BHB$\Lambda\phi$ EOS
\citep{banik:2014qja}, the LS220 EOS \citep{lattimer:1991nc}, and the
SFHo EOS \citep{steiner:2012rk}. We include an approximate treatment of
neutrino cooling using the scheme discussed in \citet{radice:2016dwd}.
Results from 29 of these simulations have already been presented in
\citet{radice:2017lry} and \citet{zappa:2017xba}. Our dataset also
contains one simulation modeling the merger of a $(1.35+1.35)\, M_\odot$
binary using the DD2 \ac{EOS} and including the effects of neutrino
absorption using the M0 scheme presented in \citet{radice:2016dwd}.
Neutrino absorption does not significantly affect the outcome of the
merger, but its inclusion is necessary for a quantitative prediction of
the \ac{EM} counterparts \citep{perego:2017wtu}. Neutrinos determine the
properties of the ejecta, and in particular their electron fraction,
especially in the polar regions \citep{sekiguchi:2015dma,
radice:2016dwd, foucart:2016rxm}. The electron fraction, in turn, is the
most important parameter determining the nucleosynthetic yields, the
nuclear heating rates, the opacities of the outflows from \ac{NS}
mergers, and consequently their optical/infrared signatures
\citep{lippuner:2015gwa}. We also performed five additional simulations
at 30\% higher resolution to check for convergence in our results.

\begin{figure}
  \includegraphics[width=\columnwidth]{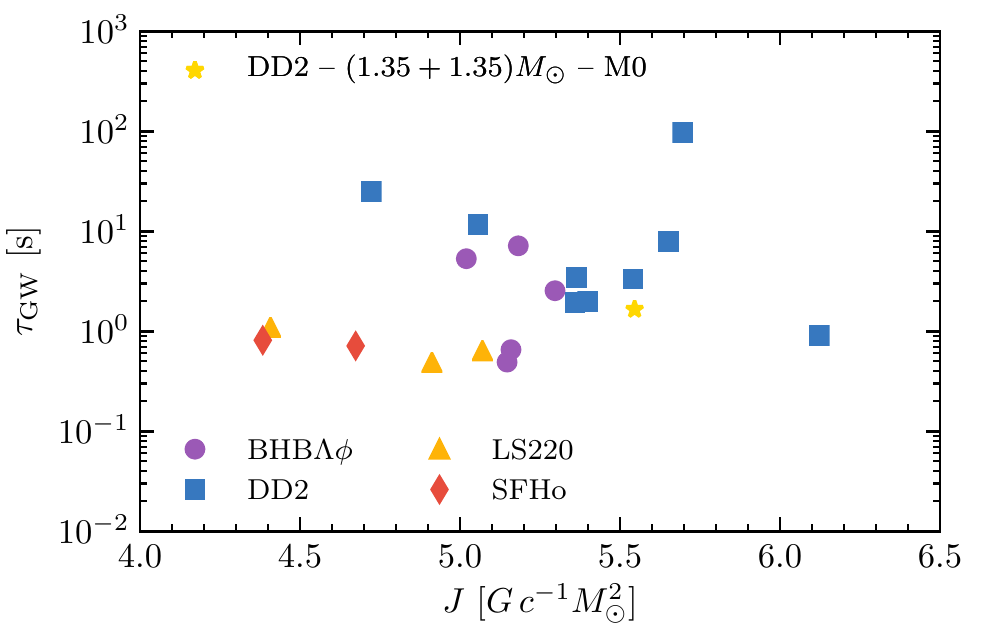}
  \caption{Gravitational wave timescale $\tau_{\rm GW} = J/\dot{J}_{\rm
  GW}$ averaged over the last millisecond of evolution for binaries
  producing massive or supramassive NS remnants. We find $\tau_{\rm GW}
  \gtrsim 0.5\, {\rm s}$, which is longer then the expected viscous
  timescale $\tau_{\rm visc} \lesssim 100\, {\rm ms}$ (see the main text).
  Note that $\tau_{\rm GW}$ grows rapidly past the initial $10{-}15\, {\rm
  ms}$ after merger, so the values reported here represent a lower
  limit.}
  \label{fig:gw.timescale}
\end{figure}

\begin{figure*}
  \begin{minipage}{0.49\textwidth}
    \includegraphics[width=\columnwidth]{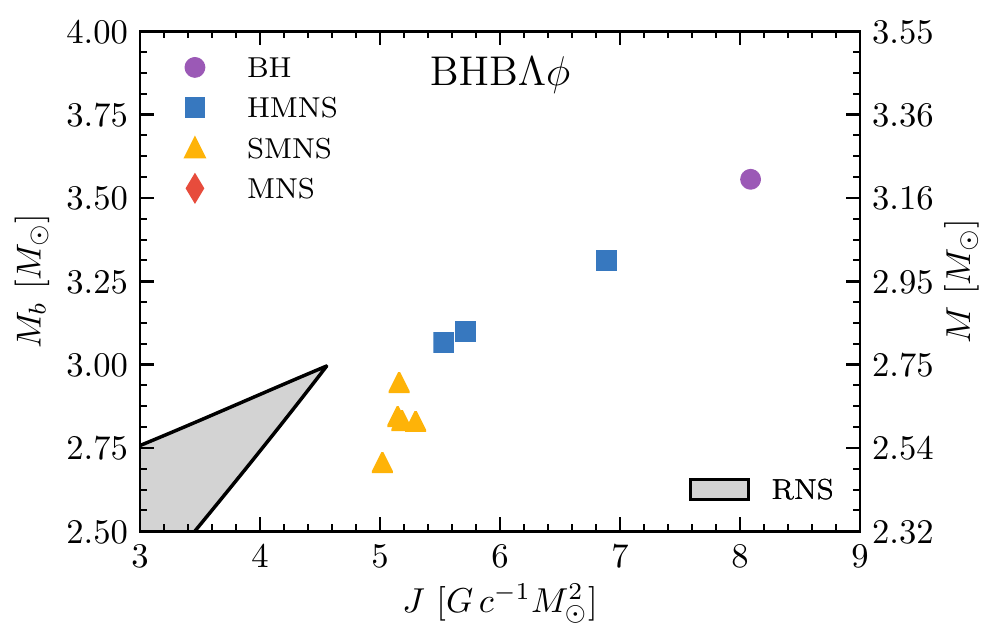}
    \includegraphics[width=\columnwidth]{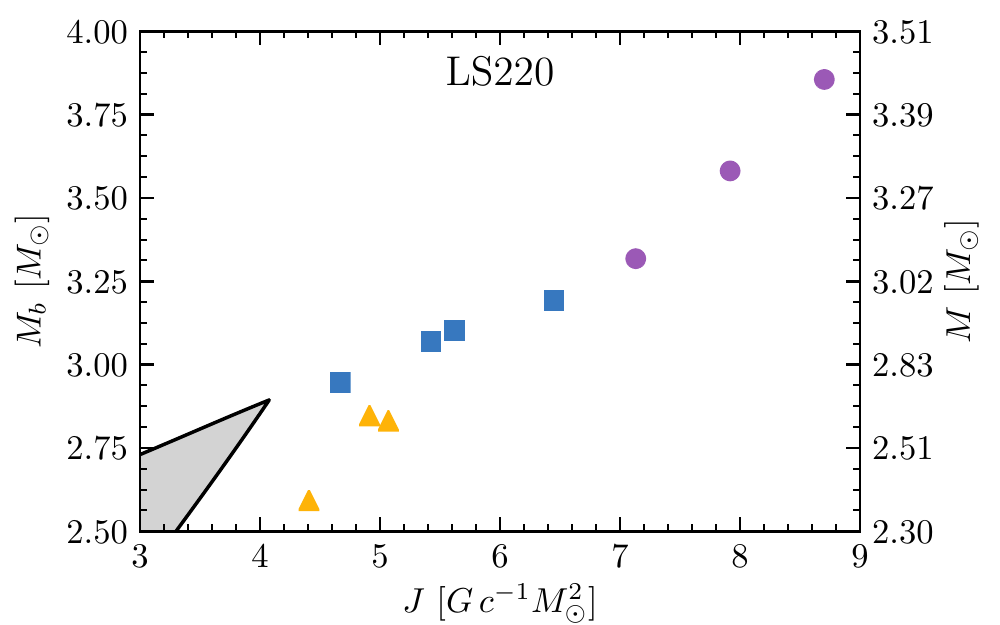}
  \end{minipage}
  \hfill
  \begin{minipage}{0.49\textwidth}
    \includegraphics[width=\columnwidth]{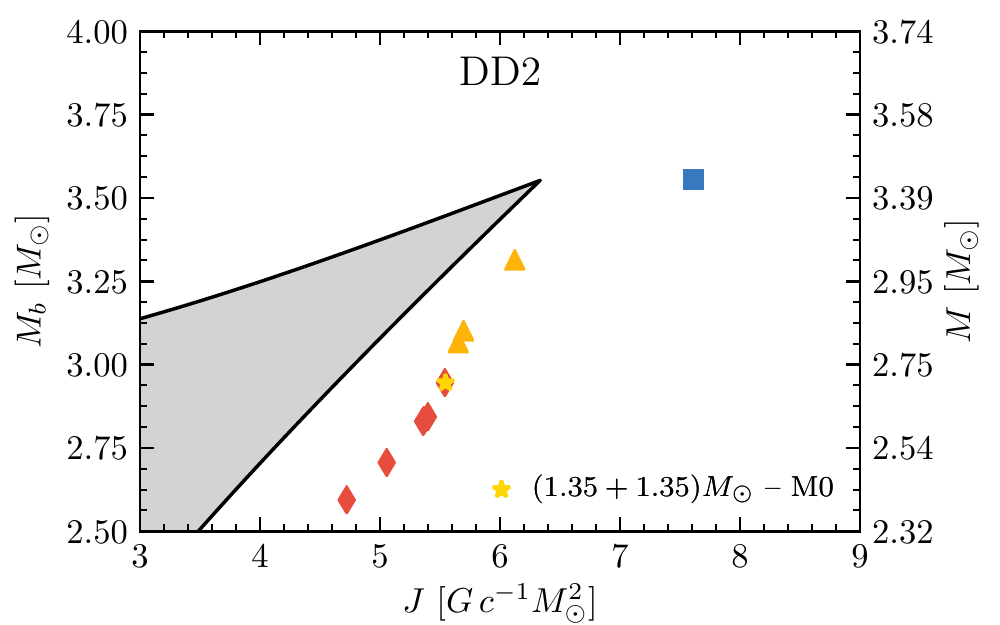}
    \includegraphics[width=\columnwidth]{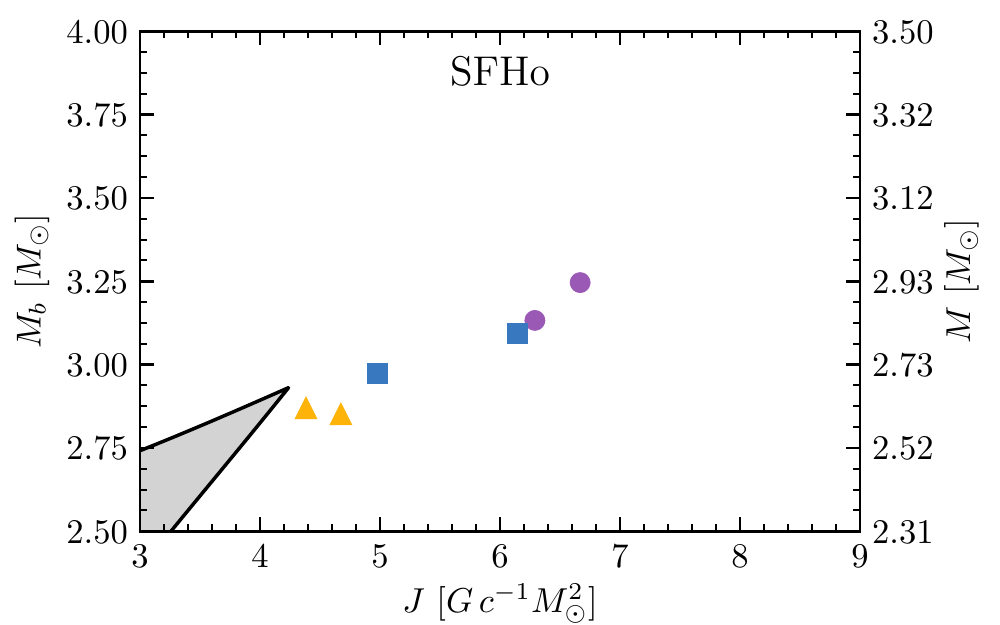}
  \end{minipage}
  \caption{Merger outcome and angular momentum at the end of the
  simulations. The grey shaded area shows the set of all rigidly-rotating
  equilibrium configurations. The gravitational mass on the right axis
  corresponds to that of an equal mass binary having the baryonic mass
  indicated by the left axis. At the end of the GW radiation timescale
  the merger remnant has significantly more angular momentum than the
  maximum allowed for rigidly rotating configurations.}
  \label{fig:final.state}
\end{figure*}

\subsection{Timescales}
We evolve each configuration for ${\sim}3{-}4$ orbits to merger and for
at least $20\, {\rm ms}$ after merger, or to \ac{BH} formation, if this
occurs earlier. We track the evolution of the total angular momentum $J$
by integrating the flux radiated by the system in \acp{GW} following
\citet{damour:2011fu} and \citet{bernuzzi:2012ci}. The integrated
$J_{\rm GW}$ is then subtracted from the angular momentum of the binary
computed by the initial data solver. We estimate the numerical
uncertainty in the determination of $J_{\rm GW}$ to be less than few
percent. Indeed, the discrepancy between standard and high-resolution
simulations is below 3\% for all of the binaries we have simulated at both
resolutions. As in previous studies, we find that gravitational angular
momentum losses in the post-merger remnant subside within
${\sim}10{-}20\, {\rm ms}$ after merger \citep{bernuzzi:2015opx,
radice:2016gym, zappa:2017xba}. At the end of our simulations the
\ac{GW} radiation timescale for angular momentum loss $\tau_{\rm GW} =
J/\dot{J}_{\rm GW}$ is typically larger than 0.5 seconds and rapidly
increasing. This is shown in Fig.~\ref{fig:gw.timescale}, where we
compute $\tau_{\rm GW}$ averaged over the last millisecond of evolution.
We want to stress that, because the \ac{GW} emission is rapidly decaying
with time, the estimate in Fig.~\ref{fig:gw.timescale} represents a
conservative lower limit. The \ac{GW} timescale should be compared to
the timescale for angular momentum transport due to turbulent viscosity.
The latter is expected to be $\tau_{\rm visc} \lesssim 100\ {\rm ms}$
\citep{hotokezaka:2013iia, kiuchi:2017zzg}. Consequently, viscosity is
the dominant mechanism determining the evolution of the remnant past the
point where we interrupt our simulations. We remark that the effective
viscosity due to small scale turbulence would further reduce the \ac{GW}
luminosity and, hence, increase the \ac{GW} timescale
\citep{radice:2017zta, shibata:2017xht}.

\subsection{Remnant Angular Momentum}
We show a summary of the final outcome of our simulations in
Fig.~\ref{fig:final.state}. As typically done in the literature, we
indicate simulations for which \ac{BH} formation occurs within one
millisecond or less after merger as ``BH''. We categorize the other
binaries according to their total baryonic mass $M_b$: if $M_b$ is
larger than the maximum baryonic mass of rigidly rotating \acp{NS}, as
predicted by the zero-temperature neutrino-less beta-equilibrated
\ac{EOS}, then the merged object is considered to be a \ac{HMNS}.
Otherwise, we distinguish between \ac{MNS} and \ac{SMNS} depending on
whether $M_b$ is smaller or larger than the maximum baryonic mass for a
nonrotating \ac{NS}, respectively. Despite the naming convection, it is
important to remark that the outcome of mergers with masses close to the
demarcation line between \ac{SMNS} and \ac{HMNS} is likely to depend on
many factors besides the maximum mass for rigidly rotating \acp{NS}. As
discussed below, mass loss, angular momentum transport, and
finite-temperature effects could all either stabilize low-mass
\acp{HMNS} or trigger an early collapse for high-mass \acp{SMNS}. For
these binaries the distinction between \acp{SMNS} and \acp{HMNS} might
not be predictive of the evolution of the remnant over timescales $t\sim
\tau_{\rm visc}$.

We use the publicly available code \texttt{RNS}
\citep{stergioulas:1994ea} to construct equilibrium sequences for
rigidly rotating \acp{NS}. The sequences are constructed assuming zero
temperature and neutrino-less beta equilibrium. For brevity, we refer to
these equilibria as being ``cold''. The gray shaded regions in
Fig.~\ref{fig:final.state} show the range they span. For a fixed $J$
lower and upper boundaries of the shaded areas are set by the mass
shedding and maximum mass limit, respectively. The tip of the shaded
region marks the maximum baryonic mass configuration supported by each
\ac{EOS} in the case of rigid rotation. Keeping in mind the caveats we
have just discussed, we label binaries with $M_b$ larger than this limit
either as \acp{HMNS} or as \acp{BH}, depending on whether a \ac{BH} was
promptly formed in the simulations or not. Our analysis shows that
\ac{MNS} and \ac{SMNS} are endowed with significantly more angular
momentum than that corresponding to the mass shedding limit for
equilibrium configurations. This can be seen from the fact that the
fast GW-drive phase of \ac{NS} mergers always ends well outside on the
right of the shaded areas in Fig.~\ref{fig:final.state}.

Our results exclude the possibility that the \acp{SMNS} formed in binary
mergers could collapse due to the lack of sufficient angular momentum
support, as proposed in \citet{ma:2017yva}. These binaries would appear
on the \emph{left} of the grey shaded area in
Fig.~\ref{fig:final.state}. Moreover, we can also exclude the
possibility that the angular momentum of \ac{SMNS} remnants could be
distributed in such a way as to leave to central part of the remnant
unstable to gravitational collapse. The reason is that the rotational
profiles of \ac{NS} merger remnants have a minimum at their center
\citep{shibata:2006nm, kastaun:2014fna, endrizzi:2016kkf,
kastaun:2016yaf, hanauske:2016gia, ciolfi:2017uak}, so the remnant's
core is expected to spin up during the viscous evolution
\citep{radice:2017zta}. Consequently, the gravitational collapse of a
hypothetical low-mass binary, if it occurs, must happen dynamically
during the merger and would have been seen in the simulations.

We find that massive or supramassive remnants need to shed excess
angular momentum before they can settle into equilibrium configurations.
The removal of angular momentum has to occur within the viscous
timescale, which is too rapid for additional \ac{GW} losses to play a
significant role. Consequently, angular momentum losses must be driven
by viscous effects and will likely be accompanied by mass losses.
Moreover, because the mass shedding limit moves to lower $J$ with
decreasing $M_b$, this process could very effectively generate large
outflows.

\begin{figure}
  \includegraphics[width=\columnwidth]{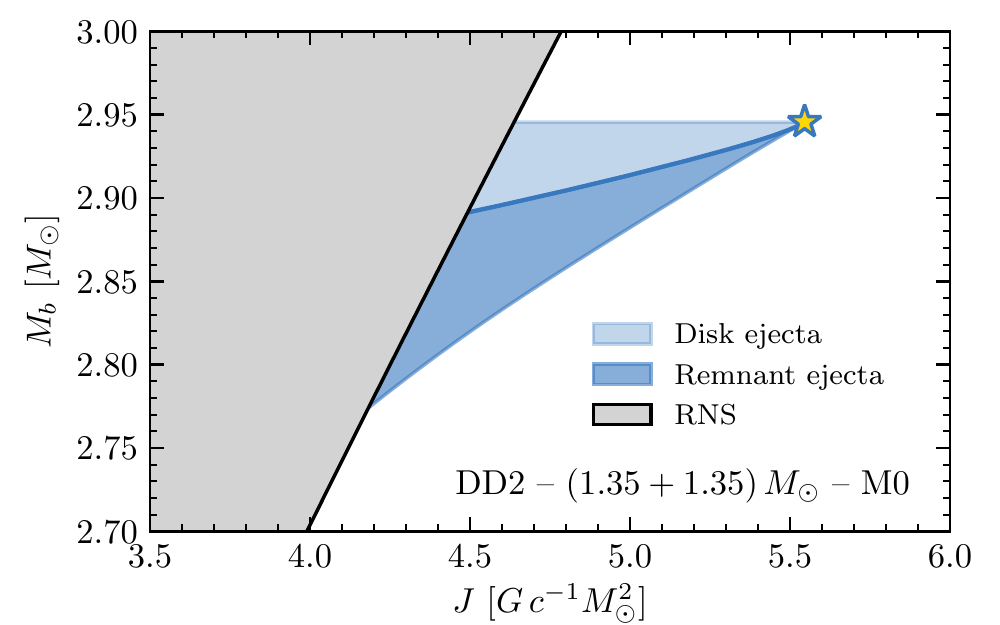}
  \caption{Estimated outcomes for the viscous evolution of a
  $(1.35+1.35) M_\odot$ binary simulated with the DD2 EOS and neutrino
  cooling/heating. The grey shaded area shows the set of all
  rigidly-rotating equilibrium configurations. The solid line is a
  conservative estimate of the mass ejection and a possible trajectory
  for the viscous evolution. The blue shaded region denotes the range of
  all possible outcomes of the viscous evolution, which we tentatively
  classify according to the underlying ejection mechanism. The first
  (disk ejecta) regime corresponds to the ejection of matter due to the
  nuclear recombination of the accretion disk. The second regime
  (remnant ejecta) is due to viscous instabilities in the merger
  remnant. Overall, we find that the merger remnant has enough angular
  momentum to unbind up to ${\sim}0.17\, M_\odot$ of material.}
  \label{fig:final.state.example}
\end{figure}

\subsection{Viscous-driven Ejecta}
We estimate an upper limit to the amount of material that could be
unbound by viscous processes after merger using 3D data taken at the end
of our simulations. We integrate the baryonic mass and the fluid angular
momentum densities\footnote{For a fluid with stress energy tensor
$T^{\mu\nu}$ this is defined as $T_{\mu\nu} n^\mu \phi^\nu$ where
$n^\mu$ is the normal to the $t = {\rm const}$ hypersurface and
$\phi^\mu$ is the generator of the rotations in the orbital plane.} on a
series of cylindrical shells. In doing so, we implicitly assume that the
spacetime is close to stationary and axisymmetric at the end of our
simulation. We find that the angular momentum of the system
estimated in this way agrees with that measured by integrating the
\ac{GW} flux to within 1\% for all models, apart from an outlier, the
LS220 binary with $(1.4+1.2)\, M_\odot$, for which the disagreement is
4\%. We start from the outer edge of the grid, and we progressively
subtract their contribution to the total mass and angular momentum. We
proceed in our integration until the resulting $M_b$ and $J$ enter the
region spanned by rigidly rotating equilibrium configurations. This
estimate is clearly an upper limit to the viscous outflow, because it
assumes that the each ejected fluid element only carries away the
angular momentum it had at the beginning. In reality, because of the
viscous angular momentum transport, the outer edge of the disk will be
endowed with some of the angular momentum initially at smaller
cylindrical radii. We remark that the main underlying assumption of our
analysis are that the minimum energy state of the system is achieved
when a uniformly rotating star is formed \citep[e.g.,][]{hartle:1967a}
and that the dynamics is dominated by the action of viscosity,
which drives the system towards this minimum energy state.

Our results are illustrated in Fig.~\ref{fig:final.state.example} for
the DD2 binary $(1.35+1.35) M_\odot$ simulated with neutrino
reabsorption, which we take as our fiducial binary. The procedure we
have just discussed generates the lower edge of the blue band in
Fig.~\ref{fig:final.state.example} representing the range of possible
outcomes for the viscous evolution. The starting point for the viscous
evolution is the end of the GW-dominated phase of the evolution -- and
the end of our simulation -- marked by the star symbol in
Fig.~\ref{fig:final.state.example}.  We find that this binary could
eject up to ${\sim} 0.17\, M_\odot$ of material during its viscous
evolution.  The upper boundary of the blue band in the figure is the
unlikely case in which angular momentum is removed without any outflow.

A more conservative estimate can be obtained assuming that the material
becomes unbound due to the recombination of nucleons into nuclei and the
subsequent liberation of nuclear binding energy, a scenario discussed in
detail in \citet{beloborodov:2008nx}, \citet{lee:2009a}, and
\citet{fernandez:2013tya}, among others. This has been shown to occur
once the material has reached a cylindrical radius $\varpi^\ast$ at
which the nuclear recombination energy equals the gravitational binding
energy \citep{lee:2009a, fernandez:2013tya}, that is
\begin{equation}
  \frac{G M m_b}{\varpi^\ast} \simeq 8.8\, {\rm MeV}\,.
\end{equation}
In the previous equation $M$ is the central object mass and $m_b$ is the
average baryon mass. Accordingly, a ring of material initially orbiting
at radius $\varpi < \varpi^\ast$ and becoming unbound would carry away,
in addition to its specific angular momentum $j(\varpi)$, also the
angular momentum needed to expand to $\varpi^\ast$. Assuming a Keplerian
disk, this implies that the angular momentum carried away by the ring
initially at $\varpi$ is
\begin{equation}\label{eq:jmodified}
  j^\ast(\varpi) = j(\varpi) \left(\frac{\varpi^\ast}{\varpi}\right)^{1/2}\,.
\end{equation}
We take $\varpi^\ast = 300\, G/c^2\, M_\odot$ as fiducial value,
corresponding to $M \simeq 2.5\, M_\odot$. We repeat the tally of
angular momentum and mass that can be removed from the remnant taking
into account the previous equation. The results are represented by the
blue line in Fig.~\ref{fig:final.state.example} laying inside the
allowed region for the viscous evolution. This yields an ejecta mass of
${\sim}0.05\, M_\odot$ for the DD2 $(1.35+1.35)\,M_\odot$ binary. Our
estimate is in good agreement with the results of
\citet{fujibayashi:2017puw}, who considered the post-merger evolution of
the same binary with 2D axisymmetric viscous GRHD simulations. They
estimated the viscous ejecta mass to be ${\sim} 0.05\, M_\odot$. Note,
however, that the mass ejection was still ongoing at the end of the
simulations presented by \citet{fujibayashi:2017puw}, so the total
ejecta mass might be larger than what they estimated.

We remark that the presence of neutrino-driven winds from the disk might
alter the dynamics with respect to the simple viscous spreading model we
have considered for our analysis. On the one hand, extant post-merger
simulations without viscosity find that the mass entrained by the
neutrino-driven wind should only be of few $10^{-3}\ M_\odot$
\citep{dessart:2008zd, perego:2014fma, martin:2015hxa,
fujibayashi:2017xsz}. So neutrino-driven winds should only amount to a
small correction to the viscous outflow. On the other hand, neutrino
heating could play an important role, together with nuclear
recombination, in unbinding material that has been transported to less
gravitationally-bound regions by viscosity \citep{lippuner:2017bfm}.
High-resolution \ac{GRMHD} studies of the evolution of post-merger
accretion disks with neutrinos will be needed to quantify the relative
importance of nuclear recombination and neutrino heating.

\begin{figure}
  \includegraphics[width=\columnwidth]{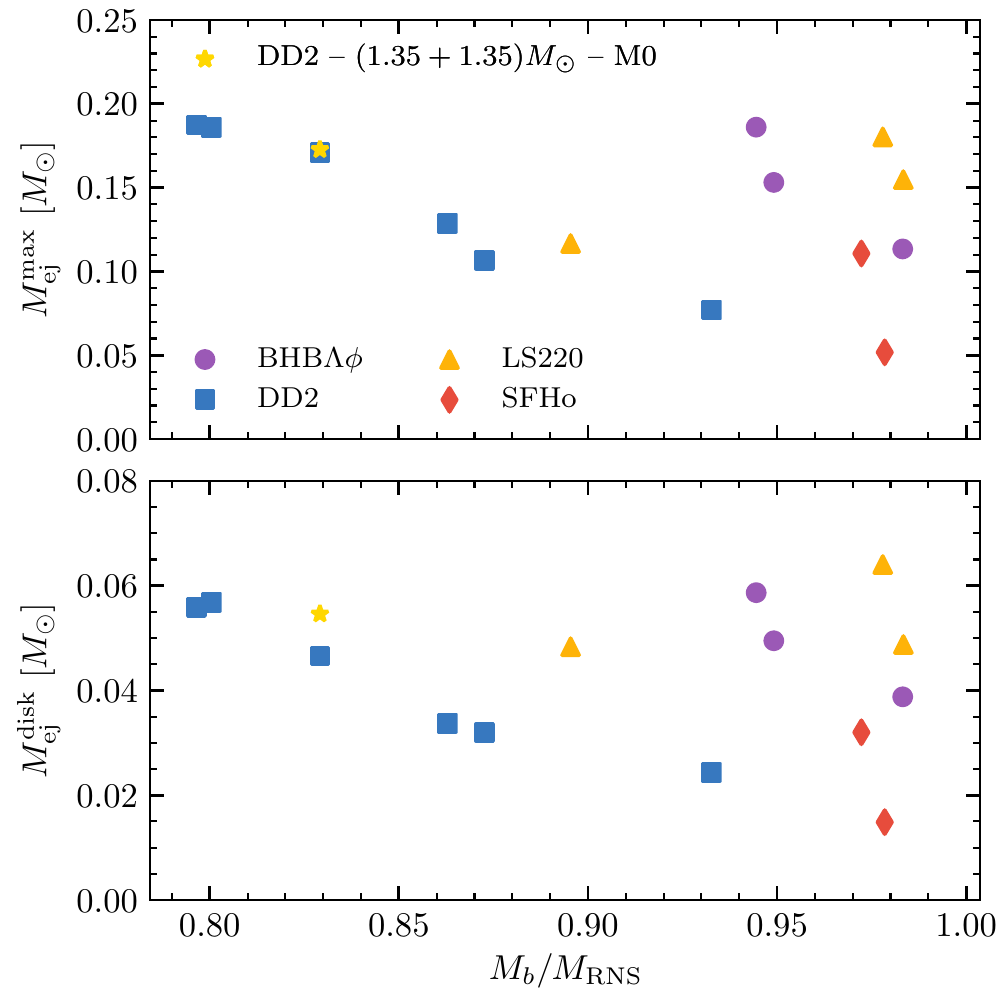}
  \caption{Upper limit of the viscous ejecta (\emph{upper panel}) and
  conservative estimate (\emph{lower panel}) as a function of the total
  baryonic mass of the binary. The masses are normalized to the maximum
  mass for uniformly rotating equilibria $M_{\rm RNS}$. Supramassive and
  massive merger remnants are expected to eject up to ${\sim} 0.2\,
  M_\odot$ of material.}
  \label{fig:ejecta}
\end{figure}

Our conservative estimate of the viscous ejecta for our fiducial DD2
$(1.35+1.35)\,M_\odot$ binary decreases by ${\sim}0.01\, M_\odot$ when
neutrino absorption is not included in the simulation (lower panel of
Fig.~\ref{fig:ejecta}). The reason is that the inclusion of neutrino
absorption inflates the outer part of the accretion disk in the region
$\varpi \gtrsim 80\, {\rm km}$. This pushes some of the material to
larger radii, where it can be unbound with a smaller expenditure of
angular momentum (Eq.~\ref{eq:jmodified}). The inner part of the remnant
is only weakly affected, so this effect is muted when computing the
upper limit on the viscous ejecta.

We point out that the evaporation of the disk due to its nuclear
recombination is not specific to binaries forming long-lived remnants.
Indeed, it is expected to occur even if the central object is a \ac{BH}
\citep{beloborodov:2008nx, metzger:2008av, lee:2009a, fernandez:2013tya,
metzger:2014ila, fernandez:2014bra, siegel:2017nub}. However, while
\acp{BH} formed in \ac{NS} mergers are well below the Kerr limit
\citep{kiuchi:2009jt, kastaun:2013mv, bernuzzi:2015opx, zappa:2017xba},
long-lived remnants necessarily have to dissipate a significant fraction
of their angular momentum within the viscous time
(Fig.~\ref{fig:final.state}). Consequently, the case of a long-lived
remnant is qualitatively and quantitatively different and could result
in more massive outflows. For this reason, we distinguish two possible
components of the viscous ejecta: the ``disk'' and the more general
``remnant'' ejecta. The first component is due to the recombination of
the disks, while the second is due to the settling of a long-lived
remnant to a uniformly rotation equilibrium. We tentatively identify the
disk ejecta component with our conservative estimate of the ejecta and
the remnant ejecta component with everything exceeding the conservative
estimate.

We repeat the analysis for 14 other binaries producing long-lived
remnants. Note that we exclude from this analysis 5 of our binaries for
which the 3D data necessary for this analysis has been lost. Our results
are shown in Fig.~\ref{fig:ejecta}. We find that the formation of
massive or supramassive \acp{NS} in binary mergers could be accompanied
by the ejection of up to ${\sim}0.2\, M_\odot$ of material within few
viscous timescales. The more conservative estimate using
Eq.~(\ref{eq:jmodified}) yields viscous ejecta mass ${\sim}4$ times
smaller. Of the five high-resolution binaries we perform to
quantify the numerical uncertainty of our simulation three form a long
lived remnant: the $(1.35+1.35)\,M_\odot$ binaries with the
BHB$\Lambda\phi$ and DD2 \ac{EOS}, and the $(1.4 + 1.2)\, M_\odot$
binary with the DD2 \ac{EOS}. The typical numerical uncertainties in the
determination of the ``disk'' and ``remnant'' ejecta are less than 25 \%
and 13\%, respectively. We conclude that ${\sim}0.05{-}0.2\, M_\odot$
of material should be generically ejected during the viscous phase of
the evolution of long-lived \ac{NS}-merger remnants.

\begin{figure}
  \includegraphics[width=\columnwidth]{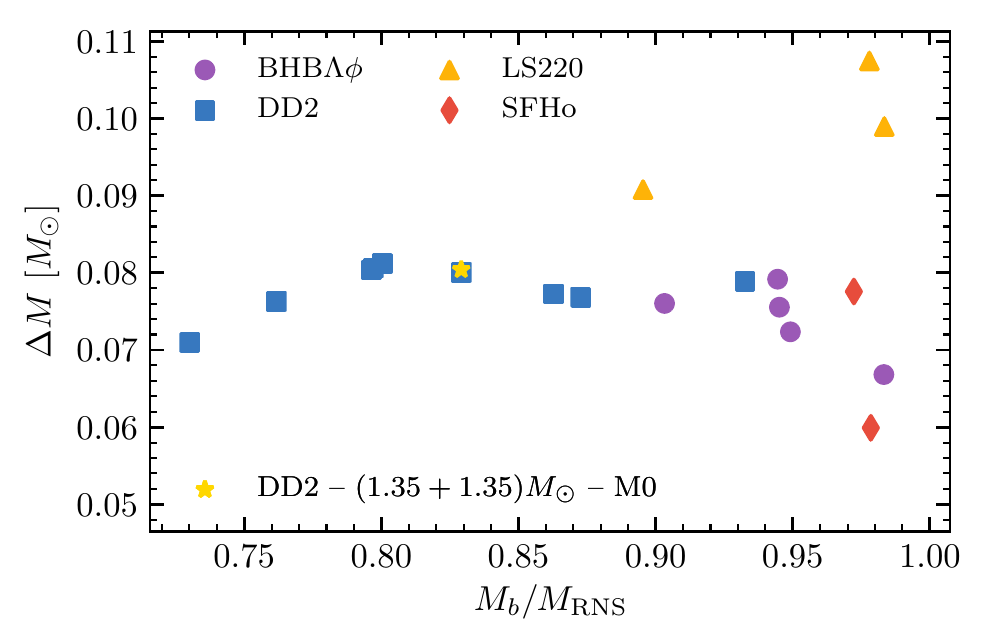}
  \caption{Difference between the gravitational mass of long-lived
  merger remnants and that corresponding to rigidly-rotating equilibrium
  configurations having the same number of baryons. Each point
  represents a simulation. Note that this estimate does not account for
  the binding energy of the material ejected by viscous driven wind.
  However, this should amount to at most a few percent correction to the
  reported values. Masses on the $x$-axis are normalized by the maximum
  mass for a rigidly rotating NSs predicted by the EOS $M_{\rm RNS}$. We
  find that long lived merger remnants need to liberate ${\sim} 0.08\,
  M_\odot$ of gravitational binding energy before settling down.}
  \label{fig:outcome.mass}
\end{figure}

\subsection{Stability of the Remnants and Neutrino Emission}
Our simulations indicate that long-lived remnants from binary \ac{NS}
mergers are not only born with excess angular momentum, but also with
excess gravitational mass compared to cold rigidly-rotating equilibria.
This is shown in Fig.~\ref{fig:outcome.mass}. We find that long-lived
\ac{NS} merger remnants have gravitational masses ${\sim}0.08\, M_\odot$
larger than the corresponding equilibrium models having the same
baryonic mass, but zero temperature. Given the long \ac{GW} timescale
and the neutrino luminosities at the end of our simulations, we can
infer that most of the excess of gravitational binding energy will be
radiated in the form of neutrinos. The cooling timescale for the massive
\ac{NS} remnant is of ${\sim}2-3$ seconds \citep{sekiguchi:2011zd}.
These conditions are analogous to those found in newly born \acp{NS} in
core-collapse supernovae (CCSNe; \citealt{burrows:1981zz,
burrows:1986me, pons:1998mm, fischer:2009af, hudepohl:2010a,
roberts:2011yw, roberts:2016rsf}).  \acused{CCSN}

Differently from \acp{CCSN}, however, the temperatures reached in
mergers are such that the maximum mass for a stable rigidly-rotating
``hot'' \ac{NS} remnant is actually smaller than that for cold
equilibria, as pointed out by \citet{kaplan:2013wra}. They found that
uniformly-rotating configurations with temperature profiles similar to
those found in simulations can support ${\sim}0.1 M_\odot$ less baryonic
mass than cold configurations. On the one hand, finite temperature and
finite neutrino chemical potential only contribute a modest ${\sim}10\%$
increase of the pressure in the core of the merger remnant, at densities
${\sim}10^{15}\, {\rm g} \, {\rm cm}^{-3}$, so finite temperature cannot
stabilize the \ac{NS} remnant against gravitational collapse. On the
other hand, thermal support inflates the mantle of the \ac{NS} remnant,
i.e., the region with subnuclear densities. Because of the extended
envelope, uniformly rotating sequences reach the mass shedding limit at
lower angular frequencies \citep{kaplan:2013wra}. This implies that a
merger \ac{NS} remnant that is formally supramassive according to the
cold \ac{EOS} could actually be hypermassive. In other words, it is
possible to form supramassive \ac{NS} remnants with baryonic masses and
thermodynamical profiles for which there is no rigidly-rotating
equilibrium. These \ac{NS} remnants could either shed their excess mass
or collapse to \ac{BH} within their viscous evolution.

\begin{figure}
  \includegraphics[width=\columnwidth]{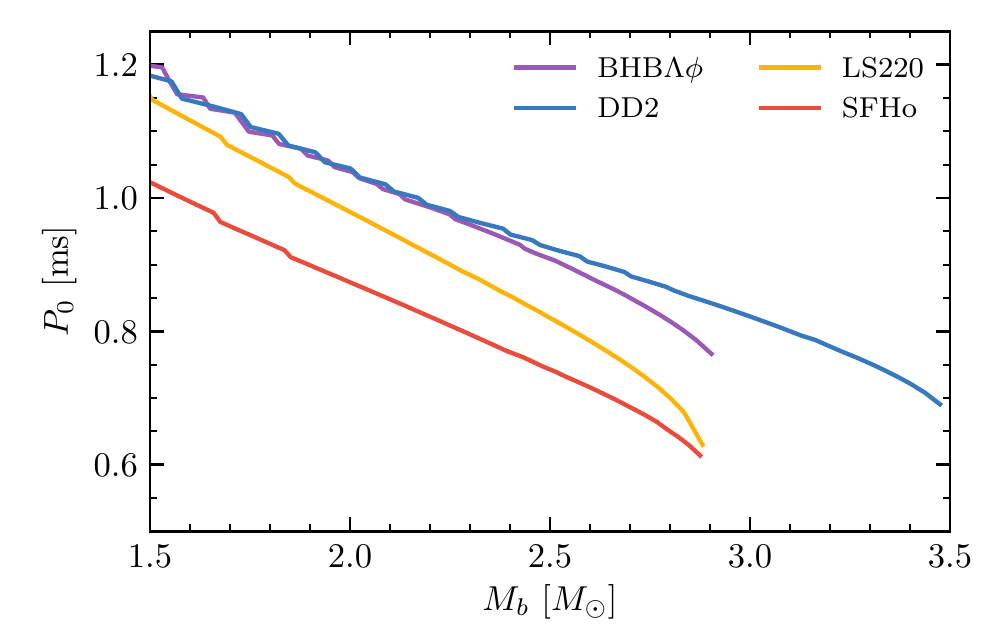}
  \caption{Rotational periods for rigidly rotating NSs at the mass
  shedding limit. This corresponds to the spin period of a long-lived
  merger remnant after viscosity has erased the differential rotation.}
  \label{fig:spin}
\end{figure}

\subsection{Spin of Long-lived NS Remnants}
Our results also imply that the outcome of the viscous evolution of
supramassive and massive NS remnants must be a rotating \ac{NS} at the
mass shedding limit with spin periods $P_0 \lesssim 1\, {\rm ms}$.  The
precise spin values can be computed using equilibrium sequences and are
shown in Fig.~\ref{fig:spin}. They depend on the baryonic mass of the
remnants at the end of their viscous evolution and can be well fitted
using a simple linear ansatz:
\begin{equation}
\label{eq:spin}
  P_0 = \left[a \left(\frac{M_b}{1\, M_\odot} - 2.5\right) + b\right]
  {\rm ms}\,.
\end{equation}
with \ac{EOS}-dependent coefficients $a\sim-(0.2{-}0.3)$ and $b\sim1$.
We report the fitting coefficients for the 4 \acp{EOS} used in this
study, as well as for other 12 representative \acp{EOS}, in
Tab.~\ref{tab:spin}. These are obtained using a standard least square
minimization in the mass interval $2.4\, M_\odot \leq M_b \leq 2.6\,
M_\odot$. The table reports also the maximum discrepancy between the
spin predicted by the fit and the actual spin as computed by RNS for
mass shedding models with $M_b > 2\, M_\odot$. We find this linear
ansatz to be an excellent approximation for binaries with total baryonic
mass larger than ${\sim}2\, M_\odot$. In particular, the maximum
relative error in the fitting interval is less than 1\%, and the maximum
error for $M_b > 2\, M_\odot$ is below $0.12$ milliseconds. The fit
slightly overestimates the spin for configurations close to the maximum
mass, especially for very soft \acp{EOS}, as can be inferred from
Fig.~\ref{fig:spin}.

Our estimated spin periods are significantly smaller than those
typically inferred for the progenitors of \ac{SGRB} with extended
emission in the context of the magnetar model. Those are typically found
to be ${\sim} 10\, {\rm ms}$ \citep{fan:2013cra, gompertz:2013aka}.  A
possible way to resolve the tension with the magnetar model would be to
assume that \ac{GW} losses could continue past the viscously-driven
phase of the evolution and spin down the remnant over a timescale of
many seconds to minutes \citep{fan:2013cra, gao:2015xle}. \ac{GW}
emission might be supported by secular instabilities in the remnant
\citep{chandrasekhar:1992pr, friedman:1978hf, lai:1994ke,
stergioulas:2003yp, corsi:2009jt, paschalidis:2015mla, east:2015vix,
doneva:2015jaa, radice:2016gym, lehner:2016wjg, east:2016zvv}, or by a
deformations due to a strong toroidal field \citep{fan:2013cra}.

\begin{table}
\caption{
Fitting coefficients $a$ and $b$ (see Eq.~\ref{eq:spin}) for the spin of
long-lived remnants at the end of the viscous evolution and maximum
error for $M_b > 2\, M_\odot$ in milliseconds $e$.
}
\begin{center}
\label{tab:spin}
\scalebox{0.9}{
\begin{tabular}{lccc}
\hline\hline
EOS & $a$ & $b$ & $e$ \\
\hline
2H & $-0.27$ & $1.18$ & $0.05$ \\
ALF2 & $-0.23$ & $0.85$ & $0.04$ \\
APR & $-0.21$ & $0.69$ & $0.12$ \\
BHBlp & $-0.27$ & $0.91$ & $0.03$ \\
DD2 & $-0.20$ & $0.93$ & $0.04$ \\
ENG & $-0.20$ & $0.77$ & $0.04$ \\
H4 & $-0.35$ & $0.94$ & $0.02$ \\
LS220 & $-0.34$ & $0.82$ & $0.06$ \\
\hline\hline
\end{tabular}
}
\hspace{-0.2em}
\scalebox{0.9}{
\begin{tabular}{lccc}
\hline\hline
EOS & $a$ & $b$ & $e$ \\
\hline
MPA1 & $-0.17$ & $0.84$ & $0.02$ \\
MS1 & $-0.21$ & $1.10$ & $0.02$ \\
MS1b & $-0.20$ & $1.07$ & $0.03$ \\
NL3 & $-0.23$ & $1.11$ & $0.03$ \\
SFHo & $-0.27$ & $0.74$ & $0.03$ \\
SLy & $-0.25$ & $0.72$ & $0.06$ \\
TM1 & $-0.31$ & $1.02$ & $0.03$ \\
TMA & $-0.35$ & $0.96$ & $0.02$ \\
\hline\hline
\end{tabular}
}
\end{center}
\end{table}

We remark that the \ac{GW} luminosity of the one-armed instability
during the first ${\sim}50\, {\rm ms}$ of the post-merger evolution is
${\sim}10^{51}\, {\rm erg}\, {\rm s}^{-1}$ and does not show strong
evidence for decay \citep{radice:2016gym}. If the one-armed instability
were to persist without damping, then it would remove all of the NS
remnant rotational energy, which is ${\sim}10^{53}\, {\rm erg}$
\citep[e.g.,][]{margalit:2017dij}, over a timescale of ${\sim}100\, {\rm
s}$. This timescale is compatible with the spin-down timescale inferred
from the magnetar model \citep{fan:2013cra}. If so, the \ac{GW} signal
from the one-armed instability would be detectable by Adv.~LIGO
\citep{theligoscientific:2014jea} and Virgo \citep{thevirgo:2014hva} up
to a distance of ${\sim}100\, {\rm Mpc}$ for optimally oriented sources
\citep{radice:2016gym}.

Alternatively, it is possible that \ac{SGRB} with extended emission
could be the result of the accretion induced collapse of white dwarfs
\citep{dessart:2008zd, abdikamalov:2009aq, bucciantini:2011kx}, although
the host environments and the offsets from the host galaxy of \acp{SGRB}
are more consistent with the expectations from \ac{NS} mergers
\citep{berger:2013jza, kumar:2014upa}.

\section{Electromagnetic Signatures}

Matter ejected during merger and the subsequent viscous evolution
synthesizes heavy elements through the so-called rapid neutron capture
process ($r$-process; \citealt{lattimer:1974a, meyer:1989a,
eichler:1989ve, freiburghaus:1999a, korobkin:2012uy, wanajo:2014wha,
just:2014fka, martin:2015hxa, lippuner:2017bfm, thielemann:2017acv,
hotokezaka:2018aui}). The resulting abundances depend sensitively on the
neutron richness (i.e., on the electron fraction $Y_{\rm e}$), entropy,
and expansion velocity of the material \citep[e.g.,][]{hoffman:1996aj,
lippuner:2015gwa, thielemann:2017acv}.  Different ejection channels
produce outflows with different properties resulting in different
nucleosynthetic yields. For the conditions relevant to \ac{NS} mergers,
the nucleosynthesis outcome depends mainly on $Y_{\rm e}$. For $Y_{\rm
e} \gtrsim 0.25$, the production of nuclei stops at mass numbers $A \sim
120$. The production of lanthanides is possible for $Y_{\rm e} \lesssim
0.25$, while even more neutron rich material ($Y_{\rm e} \lesssim 0.15$)
is necessary to synthesize actinides \citep{lippuner:2015gwa}.

\begin{figure*}
  \includegraphics[width=2\columnwidth]{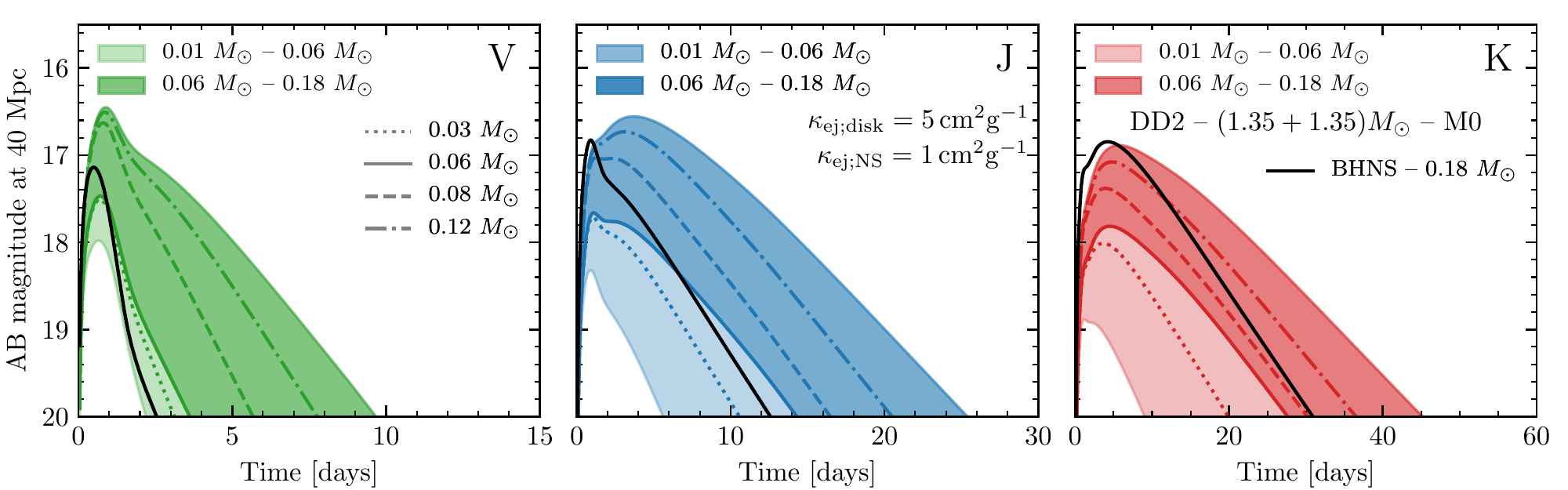}
  \caption{Kilonova color light curves for our fiducial binary (see main
  text). The colored bands correspond to the possible outcomes to the
  viscous evolution shown in Fig.~\ref{fig:final.state.example}. The
  colored solid lines correspond to the conservative estimate of the
  ejecta mass derived in Section \ref{sec: Merger Remnants}. The black
  lines are the prediction for a BHNS merger also ejecting $0.18\,
  M_\odot$ of material (see the main text for the details). The viscous
  outflows launched with the formation of long-lived NS merger remnants
  could power unusually bright kilonova lightcurves.}
  \label{fig:lightcurves}
\end{figure*}
\begin{figure*}
  \includegraphics[width=2\columnwidth]{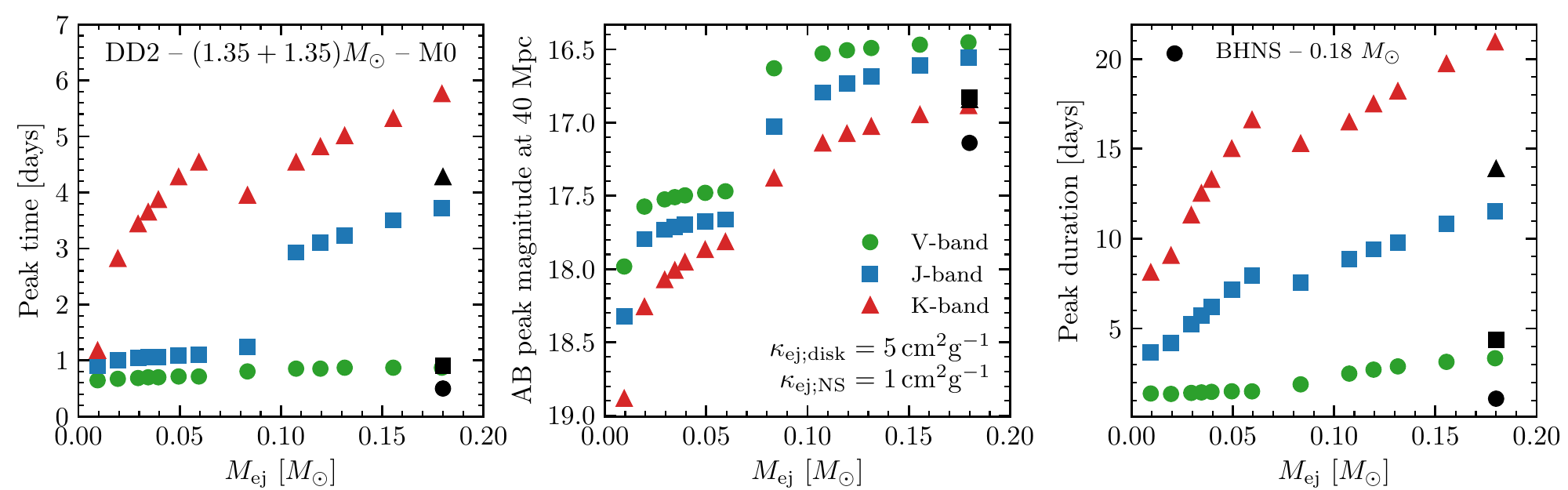}
  \includegraphics[width=\columnwidth]{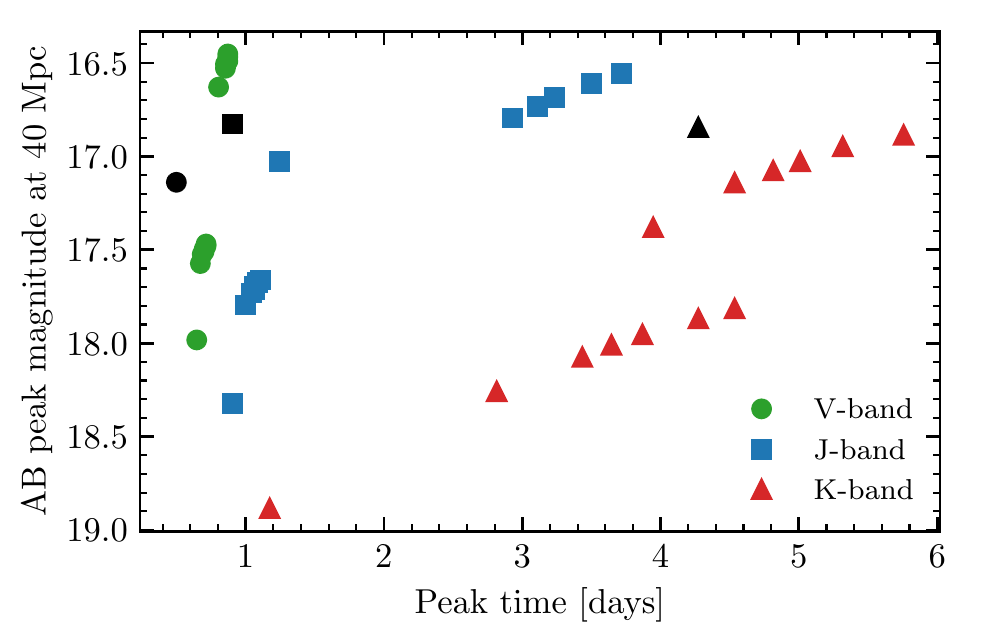}
  \hspace{0.04\columnwidth}
  \includegraphics[width=\columnwidth]{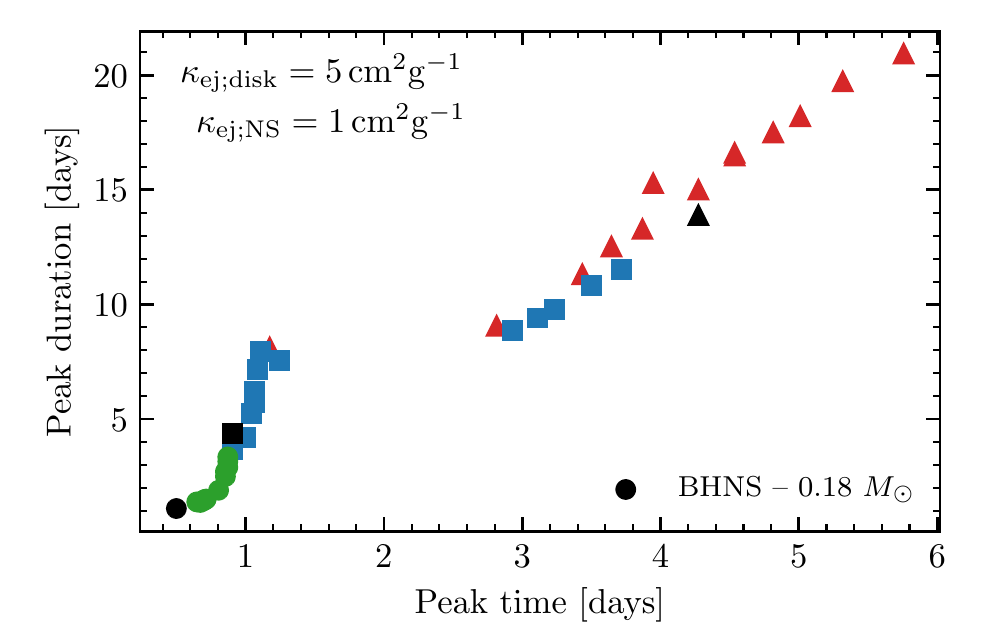}
  \caption{Kilonova peak time (\emph{upper left panel}), peak magnitude
  (\emph{upper central panel}), and peak duration (\emph{upper right
  panel}) for our fiducial binary as a function of the mass of the
  viscous ejecta. Kilonova peak magnitude vs peak time (\emph{lower left
  panel}) and peak duration vs peak time (\emph{lower right panel}). We
  find strong correlation between these key quantities and the ejecta
  mass. Note the effect of the low-opacity ($\kappa_{\rm ej;NS}=1~{\rm
  cm^2~g^{-1}}$) remnant ejecta for $M_{\rm ej}\geq 0.06 M_{\odot}$. A
  bright, slowly evolving kilonova with a blue component at early times
  would be a clear evidence for the formation of a massive or
  supramassive NS remnant in a binary \ac{NS} merger.}
  \label{fig:knprops}
\end{figure*}

The radioactive decay of the freshly synthesized $r$-process nuclei in
the ejecta powers an UV/optical/infrared transient: the kilonova
(sometimes also called macronova; \citealt{li:1998bw, kulkarni:2005jw,
metzger:2010sy, roberts:2011xz, kasen:2013xka, barnes:2013wka,
tanaka:2013ana, rosswog:2013kqa, grossman:2013lqa, rosswog:2016dhy}).
Its properties depend primarily on the rate at which radioactivity
deposits heat in the material and on the timescale over which the
expanding matter becomes transparent to thermal photons. The ejecta
composition is key to set the photon opacity of the ejecta, $\kappa$.
In particular, the presence of lanthanide and actinides is expected to
significantly increase $\kappa$, delaying the kilonova peak and shifting
its spectrum to larger wavelenghts \citep{kasen:2013xka, barnes:2013wka,
tanaka:2013ana}.

The detection of a transient compatible with a kilonova (AT2017gfo;
\citealt{arcavi:2017a, coulter:2017wya, drout:2017ijr, evans:2017mmy,
kasliwal:2017ngb, nicholl:2017ahq, smartt:2017fuw,
soares-santos:2017lru, tanvir:2017pws}) in association to GW170817
confirmed our present understanding of NS mergers and gave, for the
first time, the possibility to constraint their ejecta properties and
nucleosynthetic yields \citep{kasen:2017sxr, rosswog:2017sdn}.  The
analysis of the light curves and of the spectrum revealed the presence
of a bright, blue, component peaking at ${\sim} 1\, {\rm day}$ after the
merger, which is thought to have been powered by material moving at
${\sim} 0.3c$. This was followed by a redder component peaking at
${\sim} 5$ days and originating from more opaque and more slowly
expanding material (\citealt{chornock:2017sdf, cowperthwaite:2017dyu,
drout:2017ijr, nicholl:2017ahq, tanaka:2017qxj, tanvir:2017pws,
perego:2017wtu, villar:2017wcc}; see however \citet{waxman:2017sqv,
yu:2017syg} and \citet{li:2018hzy} for alternative interpretations).

We estimate the properties of the kilonova signature associated with the
formation of long-lived merger remnants using the semi-analytical model
we introduced in \citet{perego:2017wtu}. This includes the contribution
of ejecta with different physical origin, geometry, and thermodynamical
properties (details below). We calibrated the free parameters in this
model using AT2017gfo in \citet{perego:2017wtu}. For the calculation of
the light curves, we assume azimuthal symmetry and discretize the solid
angle in thirty slices, equally spaced in $\cos({\theta})$, $\theta$ being
the polar angle. We place the observer at a distance of 40 Mpc and at a
relative inclination of $45^{\circ}$ with respect to the symmetry axis.

We take the DD2 $(1.35+1.35)\, M_\odot$ binary with neutrino heating as
our fiducial model. We use simulation data for the dynamical ejecta,
i.e., the part of the material unbound at the time of merger, and we
vary the amount of the secular ejecta to explore the range of all
possible outcomes of the viscous evolution. For the former component, we
consider azimuthally averaged profiles of the mass, $Y_{\rm e}$, and
expansion velocity of the ejecta from the merger simulation. We assume
low effective photon opacity $\kappa_{\rm blue} = 1.0 \, {\rm cm^2
g^{-1}}$ for the ejecta with $Y_{\rm e} \geq 0.25$. We assume
lanthanide-rich opacity $\kappa_{\rm red} = 10 \, {\rm cm^2 g^{-1}}$ if
$Y_{\rm e} < 0.25$.

We also include an ejecta component due to the neutrino ablation of the
outer layers of the accretion disk. Note that this is a distinct
component of the ejecta from the viscous outflows. Following
\citet{perego:2014fma} and \citet{martin:2015hxa}, we assume that 5\% of
the disk is ejected in the form of a wind. The mass of the disk at the
end of our simulation is $0.16\, M_\odot$, so the wind amounts to
$8\times 10^{-3}\, M_\odot$ of material. Since neutrino-driven winds are
only mildly neutron rich, we assume low effective photon opacity for
this ejecta component ($\kappa_{\rm ej;wind} = \kappa_{\rm blue}$).

As discussed in the previous section, we subdivide the viscous outflow
in two parts: disk and remnant viscous ejecta. The first is assumed to
be due to the nuclear recombination of the accretion disk, and would be
present also for a short-lived remnant. The second is due to the viscous
outflow from the massive \ac{NS} and is expected only for long-lived
remnants. The disk component is expected to display a broad distribution
in $Y_{\rm e}$ which would translate in an effective opacity
intermediate between the high and low opacities of lanthanide-poor and
-rich material, respectively. Ejecta with these properties is sometimes
referred to as the purple component \citep[e.g.][]{tanaka:2017qxj,
villar:2017wcc}. For this component we take $\kappa_{\rm ej; disk} =
\kappa_{\rm purple} = 5 \, {\rm cm^2 g^{-1}}$, which is consistent with
the AT2017gfo photometry after the first few days
\citep{perego:2017wtu}. We assume the remnant viscous ejecta to be less
neutron rich than either the dynamical ejecta or the disk wind ejecta
because of the neutrino irradiation from the remnant
\citep{fujibayashi:2017puw}, and consequently we assume its opacity to
be $\kappa_{\rm ej; NS} = \kappa_{\rm blue}$. Our results do not
qualitatively change if we assume $\kappa_{\rm ej; NS} = 5 \, {\rm cm^2
g^{-1}}$, but there are quantitative differences, see
Appendix~\ref{appendix:kilonova}. We assume the disk viscous ejecta to
have a $\sin^2(\theta)$ mass distribution as in \citep{perego:2017wtu}
and the remnant viscous ejecta to be isotropic.  Expansion velocities
for both viscous outflows are taken to be spatially isotropic and with a
rms value of $0.06 c$ \citep{perego:2017wtu}. As we explore the range of
possibilities, we first switch on the disk viscous ejecta and increase
it up to a maximum value of $0.05\, M_\odot$, then we add the remnant
viscous ejecta up to the upper limit found in the previous section
$M_{\rm ej}^{\max} = 0.17\, M_\odot$.

We remark that our model does not include the thermalization of the spin
down luminosity from the merger remnant, which would further enhance the
kilonova signal \citep{yu:2013kra, metzger:2013cha, gao:2015rua,
siegel:2015swa, siegel:2015twa, kisaka:2015vma, gao:2016uwi}.  We will
explore this possibility in future works.

In Fig.~\ref{fig:lightcurves} we show light curves obtained from our
kilonova model for three representative photometric bands, namely $V$,
$J$, and $K$. The colored bands correspond to increasing values of the
ejecta mass in the viscous components. Light curves generated by varying
the amount of the disk viscous ejecta span the light shaded bands. The
light curves generated by varying the amount of the remnant viscous
ejecta span the dark shaded regions. The most relevant properties of
each light curve as a function of the total ejected mass are summarized
in Fig.~\ref{fig:knprops}. There we present the peak times, magnitudes,
and (temporal) widths of the kilonova signal. The latter are defined as
the time interval about the peak where the light curve varies by one
magnitude.

Increasing the amount of the viscous ejecta boosts the transient
brightness in all bands. However, the $V$-band peak time and duration
are only marginally affected by the presence of a large viscous ejecta.
Conversely, a large viscous ejection produces significantly brighter
peaks in the $J$ and $K$ infrared bands. The peaks are shifted to later
times and have larger temporal widths.  Notably, the increase of mass in
the remnant wind produces a second peak in the $J$ band at times longer
than one day. This peak becomes the dominant one when the remnant
viscous ejecta is turned on. The $K$ band is the most sensitive to
changes in the amount of the viscous ejecta which effect its peak
brightness, time, and duration.

\acused{BHNS}
The merger of a \ac{NS} and a \ac{BH} can also result in the dynamic
ejection of up to ${\sim}0.1\, M_\odot$ of material and in the formation
of massive accretion disks \citep{shibata:2006nm, etienne:2007jg,
duez:2008rb, etienne:2008re, pannarale:2010vs, foucart:2012nc,
foucart:2014nda, kyutoku:2015gda, foucart:2015vpa}. Extreme mass ratio
or very eccentric double \ac{NS} mergers could also eject a similarly
large amount of matter \citep{rosswog:2012wb, east:2012ww,
radice:2016dwd, dietrich:2016hky}. We investigate whether the kilonova
signal associated with the formation of a long-lived remnant in a double
\ac{NS} merger could be distinguished from the kilonova following a
\ac{BHNS} merger with a large mass ejection.

To this aim, we construct the synthetic kilonova signal for a
hypothetical \ac{BHNS} merger ejecting the same amount of material as
our fiducial binary \ac{NS} system, but with the geometry/composition
expected for \ac{BHNS} mergers. More in detail, we assume that $0.05\,
M_\odot$ of material are ejected by tidal torques. This material is
expected to be very neutron rich and have $\kappa = \kappa_{\rm red}$.
We assume that the rest of the ejecta originates from the accretion
torus formed from the tidal disruption of the \ac{NS}. Part of the disk
outflows, $0.003\, M_\odot$, are in the form of a lanthanide-free
neutrino-driven wind, for which we take $\kappa = \kappa_{\rm blue}$.
An additional $0.127\, M_\odot$ is assumed to be due to the nuclear
recombination of the disk. For the latter, we assume similar properties
to the viscous disk ejecta from \ac{NS} mergers: intermediate opacity
$\kappa = \kappa_{\rm purple}$ and $\sin^2(\theta)$ angular
distribution. The results are shown in Figs.~\ref{fig:lightcurves}
(black line) and \ref{fig:knprops} (black symbols).

We find that, while the kilonova light curves from the two systems share
some similarities, they also have important differences that would make
them distinguishable. Kilonovae associated with the formation of
long-lived remnants peak at a late time in the red bands and are
significantly brighter in all bands after the peak times.  Furthermore,
if the viscous ejecta from the remnant is lanthanide-free, as is assumed
to be the case in Figs.~\ref{fig:lightcurves} and \ref{fig:knprops},
then the kilonova peak luminosities in the blue/green bands are
significantly larger than those associated with \acp{BHNS}. On the other
hand, if the viscous ejecta from the remnant are contaminated with
lanthanides, then the peak luminosities alone are not sufficient to
distinguish long-lived remnants from \acp{BHNS}. However, the
luminosities after the peak time are still significantly larger in the
case of long-lived remnants (Fig.~\ref{fig:kn.lowopac}) that a
determination would still be possible for well observed kilonovae.

\section{Conclusions}

We have studied the outcome of \ac{NS} mergers by means of numerical
relativity simulations focusing on the properties of long-lived or
stable remnants. Our calculations employed four microphysical \ac{EOS}
and an effective treatment of neutrino cooling. We also accounted for
heating and compositional changes due to the absorption of neutrinos in
one of our simulations. We have compared the properties of long-lived
merger remnants to those of rigidly-rotating equilibrium configurations.

We have found that the post-merger starts with a short ${\sim}10{-}20\,
{\rm ms}$ phase where the evolution is mainly driven by the emission of
\acp{GW}, as also reported by \citet{bernuzzi:2015opx},
\citet{radice:2016gym}, and \citet{zappa:2017xba}. Subsequently, the
\ac{GW} luminosity drops substantially. The characteristic timescale
associated with the removal of angular momentum due to \acp{GW} exceeds
${\sim}0.5$ seconds, for some binaries by orders of magnitude, and is
still growing rapidly at the end of our simulations. This significantly
exceeds the timescale associated with the redistribution of angular
momentum operated by the effective turbulent viscosity in the remnant,
$\tau_{\rm visc} \lesssim 0.1\, {\rm s}$ \citep{hotokezaka:2013iia,
kiuchi:2017zzg}, and it is also likely to exceed the neutrino-cooling
timescale, $\tau_\nu \sim 2{-}3\, {\rm s}$ \citep{sekiguchi:2011zd}.
Thus, the remnant evolution is mainly driven by the effects of viscosity
and neutrino losses. After having reached solid body rotation and over
even longer timescales of many seconds, minutes, or hours, the remnant
spins down due to residual \ac{GW} losses and magnetic torques.

The evolution of the remnants over the viscous time is non trivial. The
reason is that, after the short, GW-driven, post-merger transient, the
\ac{NS} merger remnants are still endowed with too much angular momentum
to reach an equilibrium. More precisely, we have shown that there exists
no uniformly-rotating equilibrium configuration to which the merger
remnant can relax under the action of viscosity while conserving baryon
mass and angular momentum. Instead, massive and supramassive \acp{NS}
formed in mergers need to dissipate a significant fraction of their
angular momentum within the viscous timescale. Angular momentum
redistribution is likely be accompanied by the emission of massive
outflows since \ac{GW} losses are negligible during this phase of the
evolution. These viscous-driven outflows could potentially exceed those
typically expected from neutrino-driven winds and from the nuclear
recombination of the remnants' accretion disk. Our results indicate
that, for a typical binary, the transition to a uniformly rotating
equilibrium could be accompanied by the ejection of up to ${\sim}0.2\,
M_\odot$ of material. The mass ejection is expected to be driven
by a combination of effective turbulent viscosity, nuclear
recombination, and neutrino heating. However, the details of the
ejection process are still not well understood, especially when
long-lived remnants are formed. Long-term high-resolution
neutrino-radiation \ac{GRMHD} simulations will be needed to understand
how post-merger disks evolve.

Massive and supramassive merger remnants have gravitational masses
${\sim}0.08\, M_\odot$ larger than those of equilibrium configurations
having the same number of baryons. Our results suggest that most of the
associated energy is liberated with the emission of neutrinos on a
cooling timescale of few seconds. At the same time
\citet{kaplan:2013wra} showed that, for the temperatures reached in
mergers, trapped neutrinos and thermal support yield only minor
contributions to the pressure in the core of the remnant. However, hot
rigidly-rotating equilibrium sequences with increasing angular frequency
reach the mass shedding limit before cold beta-equilibrated sequences.
Consequently, the maximum baryonic mass achievable for hot
rigidly-rotating \acp{NS} is ${\sim}0.1\, M_\odot$ smaller than that of
cold rotating \acp{NS}.  We deduce that the fate of binaries with total
masses close to the threshold for the formation of \acp{HMNS} depends on
a complex interplay between mass ejection and neutrino cooling whose
outcome is difficult to predict. For example, remnants with masses below
the maximum for cold rigidly-rotating \acp{NS} could still collapse
because of the gravitational mass excess with which they are formed.
Conversely, massive remnants could become stable following the ejection
of large amounts of material during their viscous evolution.
Understanding the long-term evolution of systems with masses close to
this threshold is urgent, especially in view of the current efforts to
constrain the \ac{NS} \ac{EOS} using the outcome of \ac{NS} mergers
\citep{margalit:2017dij, rezzolla:2017aly, ruiz:2017due}. This will be
object of our future work.

Even though our models cannot yet predict the precise path undertaken by
the viscous evolution of the remnant, we can nevertheless constrain the
spin of the remnant once solid-body rotation has been established. This
is because, according to our simulations, the end result of the viscous
evolution must be close to the mass-shedding limit. This corresponds to
spin periods $P_0 \lesssim 1\, {\rm ms}$. We have shown that these can
be estimated from the final baryonic mass of the remnant using a simple
linear fit. The values we found are, however, much smaller than those,
around $10\, {\rm ms}$, typically inferred from the analysis of
\acp{SGRB} in the context of the magnetar model \citep{fan:2013cra,
gompertz:2013aka}. This tension could be resolved under the assumption
that \ac{GW} losses persist even after the remnant has reached solid
body rotation. The spin down timescale associated with this persistent
emission could be $\tau_{\rm GW} \sim 100\, {\rm s}$ \citep{fan:2013cra,
gao:2015xle}. \ac{GW} observations of a nearby merger event forming a
long-lived remnant might detect this extended signal or severely
constrain the magnetar model\footnote{See also \citet{bartos:2012vd} and
\citet{fan:2017rkg} for other possible applications of \ac{GW} astronomy
to the study of \acp{SGRB}.} \citep{fan:2013cra, gao:2015xle}.

We have used the model of \citet{perego:2017wtu} to produce synthetic
lightcurves of kilonovae associated with the formation of long-lived
\ac{NS} merger remnants. We have found that the inclusion of
viscous-driven ejecta from the merger remnant, in addition to the other
outflow components, can boost the peak brightness of the emission by up
to one magnitude in all bands. It also significantly broadens the width
of the light curves and shifts the peak time in the near infrared by up
to several days. The resulting kilonova is peculiarly bright, blue, and
slowly evolving, and would be easily distinguished from kilonovae
associated with \ac{NS} mergers producing \acp{BH} or \ac{BHNS} mergers,
despite the fact that the formers can also produce large outflows. Its
detection in concomitance with a \ac{SGRB} or a \ac{GW} event would
constitute smoking gun evidence for the formation of a long-lived
remnant.

\section*{acknowledgments}
It is a pleasure to acknowledge J.~Roulet for help with the \texttt{RNS}
code, W.~Del~Pozzo for help with optimizing and improving the kilonova
code, and A.~Burrows, K.~Hotokezaka, and K.~Murase for discussions.
DR acknowledges support from a Frank and Peggy Taplin Membership at the
Institute for Advanced Study and the Max-Planck/Princeton Center (MPPC)
for Plasma Physics (NSF PHY-1523261).
AP acknowledges support from the INFN initiative "High Performance
data Network" funded by CIPE.
DR and AP acknowledge support from the Institute for Nuclear Theory
(17-2b program).
SB acknowledges support by the EU H2020 under ERC Starting Grant,
no.~BinGraSp-714626.
BZ acknowledges NASA NNX15AK85G for support.
Computations were performed on the supercomputers Bridges, Comet, and
Stampede (NSF XSEDE allocation TG-PHY160025), on NSF/NCSA Blue Waters
(NSF PRAC ACI-1440083), Marconi (PRACE proposal 2016153522), and
PizDaint/CSCS (ID 667).

\bibliographystyle{mnras.bst}
\bibliography{references}

\appendix

\section{Effect of Remnant Ejecta Opacity}
\label{appendix:kilonova}

\begin{figure*}
  \includegraphics[width=2\columnwidth]{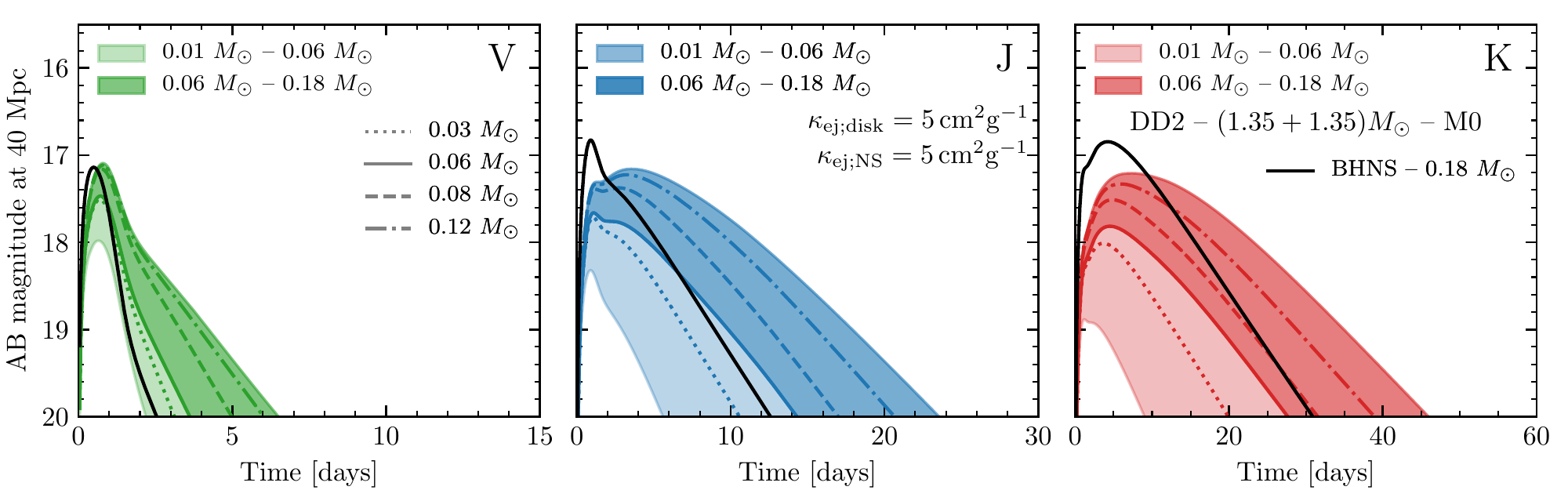}
  \includegraphics[width=2\columnwidth]{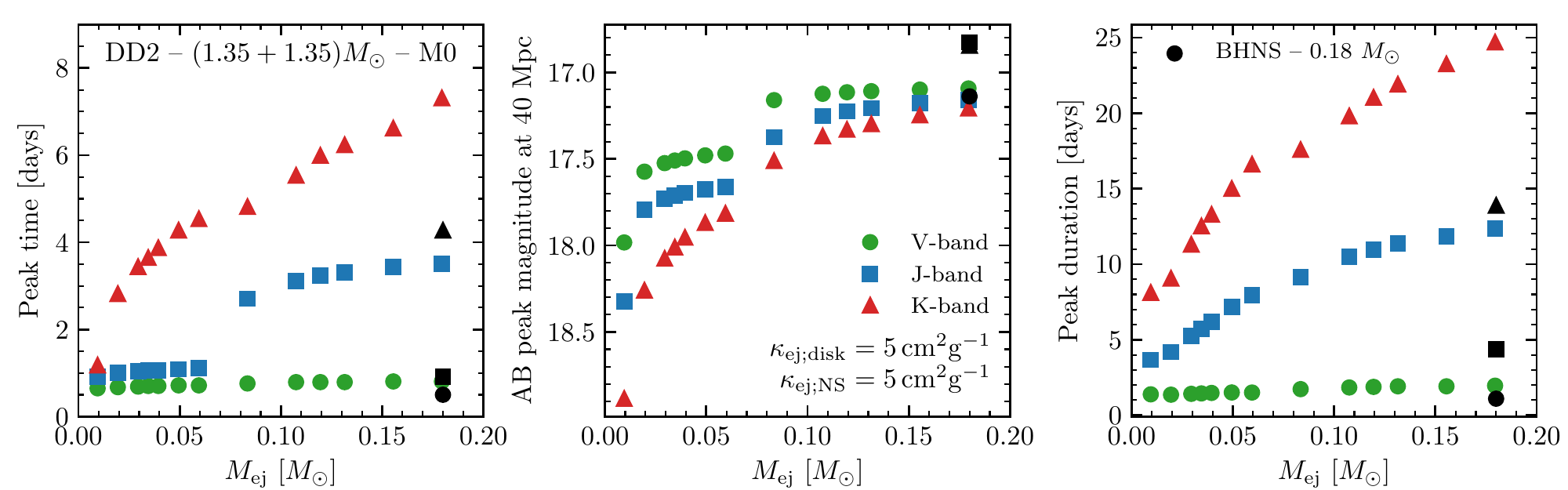}
  \includegraphics[width=\columnwidth]{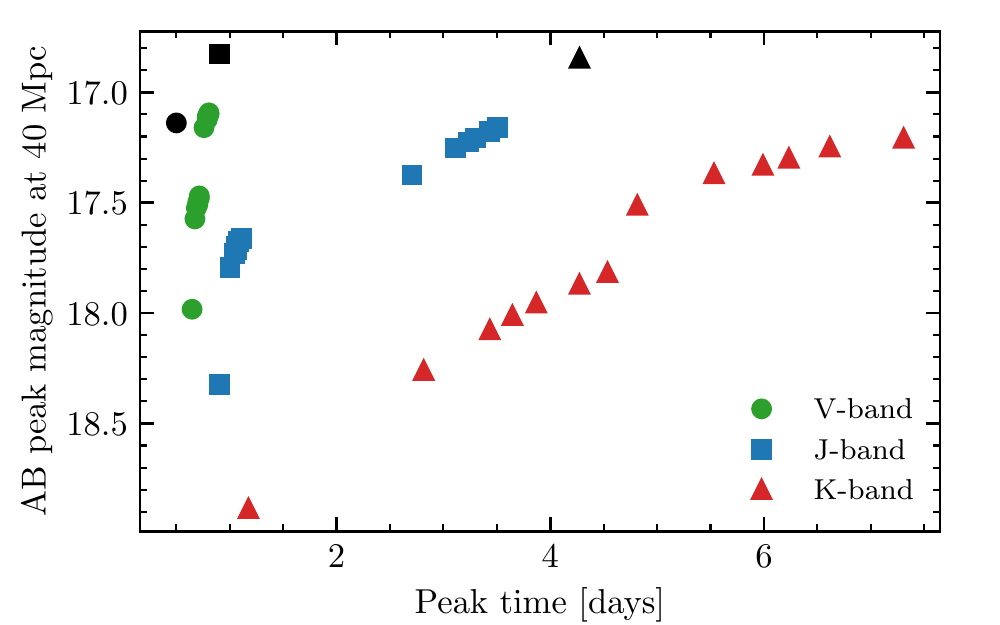}
  \hspace{0.04\columnwidth}
  \includegraphics[width=\columnwidth]{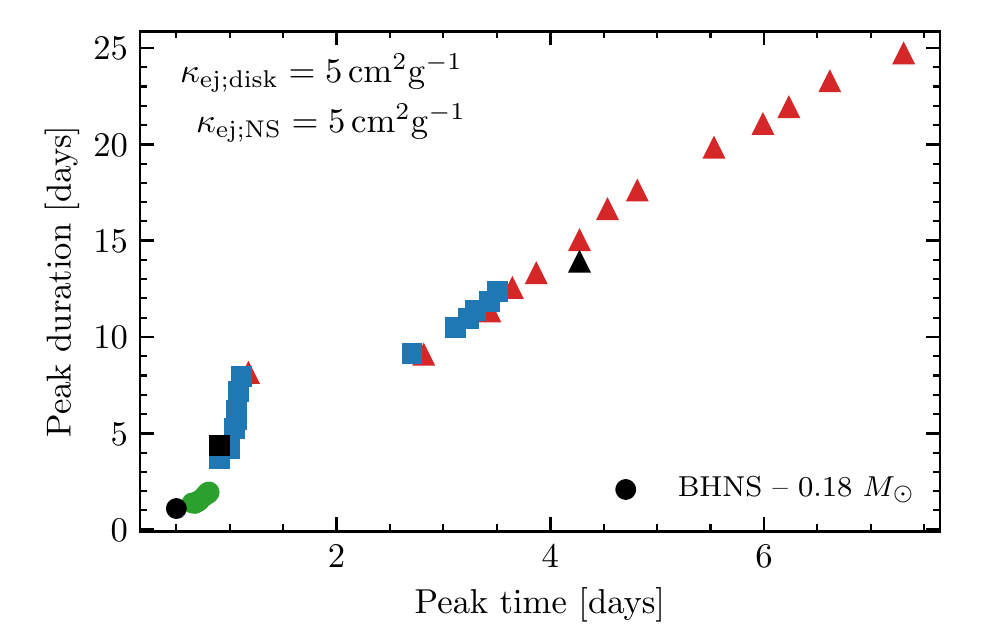}
  \caption{Kilonova light curves (\emph{top panel}), and dependency of
  the kilonova peak properties on the ejecta mass (\emph{lower panels})
  for our fiducial binary. Here, we assume the additional ejecta
  component from the SMNS to be contaminated with lanthanides, with an
  opacity of $\kappa_{\rm ej; NS} = 5\, {\rm cm}^2 {\rm g}^{-1}$. This
  figure should be contrasted with Figs.~\ref{fig:lightcurves} and
  \ref{fig:knprops} which are generated assuming $\kappa_{\rm ej; NS} =
  1\, {\rm cm}^2 {\rm g}^{-1}$.}
  \label{fig:kn.lowopac}
\end{figure*}

\end{document}